\documentclass{ws-procs975x65}

\begin{document}

\title{Microlensing in globular clusters: the first confirmed lens}

\author{Philippe Jetzer}

\address{
Institute of Theoretical Physics\\
           University of Z\"urich, 
Winterthurerstarsse 190,CH-8057 Z\"urich, Switzerland\\
E-mail: jetzer@physik.uzh.ch}


\abstract{
Microlensing observations toward globular clusters could be very useful
to probe their low mass star and brown dwarf content. 
Using the large set of microlensing events detected
so far toward the Galactic centre we investigated whether for some
of the observed events  
the lenses are located in the Galactic globular clusters.
Indeed, we found that in four cases some events might be due to lenses
located in the globular clusters themselves.
Moreover, we discuss a microlensing 
event found in M22.
Using the adaptive optics system NACO at ESO VLT it was possible to
identify the lens, which turned
out to be a low mass star of about 0.18 solar masses in the globular cluster M22
itself.}

\section{Introduction}

A sizable fraction of the mass of globular clusters could 
be in form of brown dwarfs and low mass stars. This is still an open issue and a possible way to test 
this is to use microlensing observations, see Refs.~\citen{pacz94,luca,straessle}. The idea is to monitor
Galactic Globular Clusters (hereafter GGCs)
in front of rich background fields of stars
of the galactic bulge. 
In this case, when the lens belongs to the cluster population,
its distance and velocity are roughly known. This way it is possible to get 
a more accurate estimate for the lens mass. 
Such a study has already been performed ~\cite{luca,straessle} and some events were found
which might be associated with lenses in globular clusters, see Refs.~\citen{straessle,sahu,luca,pawel}. 
 
We analysed the possible microlensing events in  
a large set of GGCs some of which are highly aligned 
with a non negligible number of microlensing events 
detected toward the Galactic Centre (hereafter GC).
The data set included 4697 microlensing events
detected in the last years by the MACHO, OGLE,
and MOA collaborations in direction of the GC.
In our analysis we focused on the configuration in which the lens is hosted in a GGC
and the source is located either in the
Galactic disc or bulge.

For a microlensing event which occurred in the year 2000 toward the
cluster M22, observed against the dense stellar field of the Milky Way
Bulge, we made a new observation in 2011.
Using the adaptive optics system NACO at ESO VLT it was possible to
identify the lens, which turned
out to be a low mass star of about 0.18 solar masses in the globular cluster
itself \cite{pawel}. 

\section{Results} 

In order to discriminate among events due to lenses hosted either in GGCs
or in the Galactic bulge/disc, 
we first made a rough selection of events being aligned with a GGC.
In particular, for every given GGC, we considered a sphere of radius $r_t$
(corresponding to its tidal radius),
centred at the GGC centre,
and we selected, as a first step, only the events being included in one such contour.
By doing so, out of the considered 4697 events, we were left with 118 only.

Due to the GGC structure, we expect the predicted number of events to
be largest toward
their centres and to decrease as we move toward their borders.
Since the alignment between an event and a projected cluster contour 
does not assure that the deflector belongs to
the GGC, this alignment possibly being accidental,
we made a further, rough selection
and considered only the events being included in the projected 
contour of a sphere centred at a GGC centre and of radius $r_i=2\times r_t/5$
(this including on average $90 \%$ of the total cluster mass).
We then distinguished between $inner$ and $outer$ events, the former being inside $r_i$
and the latter being included in the circular ring of internal radius $r_i$ and outer radius $r_t$.
Furthermore, we assumed all the outer events to be due to Galactic 
bulge/disc deflectors (this possibly underestimating the events due to GGC lenses),
whereas we left open the possibility that among the inner events
some could still be attributed to bulge/disc deflectors.
At the end we are left with 28 inner events,
among which 7 (17/4) have been detected by the MACHO (OGLE/MOA) collaboration.

\begin{table}[h]
\tbl{GGCs with inner events.
For each of them $N_{in}$ is the number of events inside a projected radius 
$r=2\times r_t/5$
and, for this subset of aligned events,
$<t_E>$ is the mean Einstein time (in days)
and $<m>$ is the average predicted lens mass in units of
solar masses.
$N_{GGC}$ ($N_{BD}$) is the number of events, out of $N_{in}$, that we expect to be due to GGC (Galactic bulge/disc) lenses.
$\Gamma_{exp}$ is the expected event rate in units of $f\times \mu_o^{-1/2}\times 10^{-3}/year$.
\label{GGC2}}
{\begin{tabular}{c|cccccc}
\hline             
{Cluster ID} &$N_{in}$& $<t_E>$  & $<m>$  &$N_{BD}$    & $N_{GGC}$ &$\Gamma_{exp}$ \\
\hline                                                          
 NGC 6522     &   8   &   13.1   &  1.63  & 4.1$\pm$ 2.0& 3.9      &  0.66       \\
 NGC 6528     &   7   &   13.0   &  2.98  & 4.9$\pm$ 2.2& 2.1      &  0.09        \\
 NGC 6540     &   7   &   17.2   &  0.06  & 4.2$\pm$ 2.0& 2.8      &  1.56      \\
 NGC 6553     &   4   &   35.7   &  0.62  & 0.6$\pm$ 0.8& 3.4      &  0.08       \\

\hline

\end{tabular}}
\end{table}

An estimate of the predicted 
number of events, $N_{GGC}$, due to low mass stars in a given GGC,
can be roughly made as follows.
Assuming that all the outer events are due to Galactic bulge/disc lenses,
we calculate how many such events, $N_{BD}$, are expected in the inner region of a 
GGC contour assuming that the number of events 
is proportional to the covered area
and that the background source distribution is uniform inside every GGC contour.
Thus we assume that the microlensing rate for Galactic bulge/disc events
is constant over the entire small
area within the tidal radius of the considered globular cluster.
Thus $N_{BD}$ is simply proportional
to the monitored area. Clearly, also with these assumptions, which are reasonable,
given the very small area considered, one expects fluctuations in the number
of events in a given area. We assume the
fluctuations to follow Poisson statistics, in which case they are given by
$\sim \sqrt{N_{BD}}$.
By doing so, for every GGC considered,
$N_{GGC}$ turns out to be around 2-4 per cluster (see Table \ref{GGC2}) and in two cases (NGC 6522 and NGC 6553)
this number is larger than the estimated fluctuation of $N_{BD}$.
Given these numbers we cannot claim for any clear evidence of
lenses hosted in GGCs. Nonetheless, it is remarkable that for 
the 4 cases considered the value of $N_{GGC}$ is positive
and most probably underestimated, since the assumption that all the events 
lying in the outer ring are due to bulge/disc deflectors possibly overestimates $N_{BD}$.

Assuming that the deflector is a GGC low mass star or brown dwarf, we can estimate its mass
through the relation $R_E/t_E=v_r$, where $v_r$ is the 
lens-source relative velocity orthogonal to the l.o.s.,
$t_E$ is the event Einstein time and $R_E$ 
is the Einstein radius. For $v_r$ we adopt the value of the proper motion of the considered
globular cluster as given in the literature.
As reported in Ref.~\citen{harris}, the mean GGC tidal radius is of the order
of tens of pc,  this making the GGC extension relatively small compared to 
the average lens distance from Earth or the source distance 
(of the order of kpc),
since we are assuming Galactic bulge/disc sources and the GGCs are kpcs 
away from the Sun.
For this reason, we make the simple assumption that in 
a given GGC the objects acting as lenses are all at the 
same distance from the Sun.
Table \ref{GGC2} shows, for the whole subset of inner events, the predicted deflector mass in units of solar masses, $<m>$, 
obtained with these assumptions.
The resulting average lens mass are values in the range $\{10^{-2},10\}$, suggesting that the involved deflectors are possibly 
either brown dwarfs, M-stars or stellar remnants. Moreover, Jupiter-like deflectors are not definitively excluded, since, 
already a small increase on $D_{os}$ can substantially reduce the predicted lens mass.

The average expected lens mass has been drawn from the set of inner events,
some of which being possibly not due objects located in the GGC.
This source of contamination should be removed before
one makes any prediction, but since we are not
able to do such a distinction, the average values on the overall inner sample can be
taken as a first crude approximation.

Also given in Table \ref{GGC2} is the number of expected events toward
the GGC centres, $\Gamma_{exp}$, as calculated through formula (36) of Ref~\citen{straessle},
where it is assumed that all the lenses have the same mass $\mu_o$,
in units of solar masses, and that their distribution is
very narrow with respect to that of the source population. $\Gamma_{exp}$ is given in units 
of $f\times \mu_o^{-1/2}\times 10^{-3}/year$,
$f$ being the fraction of matter in form of brown dwarfs, dim stars or stellar remnants in the cluster.
For a typical value of $10^2-10^3$ monitored source stars behind a GGC (this number depending also on the GGC extension)
and an observation period of $\sim$ 5 to 10 years, we expect at most between an event and a couple
of events toward each GGC depending also on the value of $f$, in reasonable
agreement with the number of events $N_{GGC}$ reported in the Table \ref{GGC2}.

In 2000 July/August a microlensing event occurred at a distance of 2.33 arc minutes
from the centre of the globular cluster M22 (NGC 6656), observed against the
dense stellar field of the Milky Way bulge. 
In order to check the hypothesis that the lens belongs to the globular
cluster we made a dedicated observation,
using the adaptive optics
system NACO at the ESO Very Large Telescope to resolve the two objects - the lens and the source - that
participated in the event. The position of the objects
measured in July 2011 was in agreement with the observed relative proper motion
of M22 with respect to the background bulge stars. Based on the brightness of
the microlens components we found that the source is a solar-type star located at
a distance of 6.0 $\pm$ 1.5 kpc in the bulge, while the lens is a
0.18 $\pm$ 0.01 $M_{\odot}$ dwarf
member of the globular cluster M22 located at the 
known distance of 3.2 $\pm$ 0.2 kpc from the Sun \cite{pawel},
therefore being the first confirmed microlens in a globular cluster. 

\section{Conclusions} 

Given these very promising preliminary results, nicely supported by 
the first confirmed event, 
it would be desirable to get more observations of globular clusters.
Indeed  few events 
could already help very much to constrain the low mass star and brown dwarf content 
and thus to get a clear mass budget and information on the stellar mass function of globular clusters.
Clearly, the expected
number of events, and thus the rate, is certainly quite small,
which would require to develop an appropriate strategy for a 
systematic survey during many years of 
the lines of sight comprising several globular clusters, in particular
the ones we analysed and for which we found some evidence of possible candidates and
also toward M22, which is a rather massive globular cluster
for which we expect  
a higher event rate as compared to the other clusters.

\end{document}